\documentclass[pra,twocolumn,superscriptaddress,amsmath,amssymb,aps,10pt]{revtex4-1}

\usepackage{graphicx}
\usepackage{dcolumn}
\usepackage{bm}
\usepackage{braket}
\usepackage{hyperref}
\hypersetup{
    colorlinks=true,        
    linkcolor=blue,          
    citecolor=blue,        
    filecolor=magenta,      
    urlcolor=blue           
}

\usepackage{color} 				
\usepackage[normalem]{ulem} 	
\definecolor{myblue}{rgb}{0.4, 0.3, 0.7}
\definecolor{newcolor}{RGB}{32, 178, 70}

\begin{document}

\title{
Thermodynamic characteristics of ideal quantum gases in harmonic potentials
within exact and semiclassical approaches}

\author{Valeriia Bilokon}
\author{Elvira Bilokon}

\author{Alexander Peletminskii}
\author{Andrii Sotnikov}
\email{a\_sotnikov@kipt.kharkov.ua}
\affiliation{Karazin Kharkiv National University, Svobody Sq. 4, 61022 Kharkiv, Ukraine}
\affiliation{Akhiezer Institute for Theoretical Physics, NSC KIPT, Akademichna Str.~1, 61108 Kharkiv, Ukraine}

\date{\today}

\begin{abstract}
We theoretically examine equilibrium properties of the harmonically trapped ideal Bose and Fermi gases in the quantum degeneracy regime.
We analyze thermodynamic characteristics of gases with a finite number of atoms by means of the known semiclassical approach and perform comparison with exact numerical results. 
For a Fermi gas, we demonstrate deviations in the Fermi energy values originating from a discrete level structure and show that these are observable only for a small number of particles.
For a Bose gas, we observe characteristic softening of phase transition features, which contrasts to the semiclassical predictions and related approximations.
We provide a more accurate methodology of determining corrections to the critical temperature due to finite number of particles.
\end{abstract}

\maketitle

\section{Introduction}
Cold atomic gases in external harmonic traps are the most prototypical examples of tunable many-body systems in the regime of quantum degeneracy. 
Over the last three decades, the experimental cooling and trapping techniques became well established and many observations are supported by convincing theoretical analysis~\cite{Dalfovo1999,Giorgini2008}.

At the same time, while learning the subject of non-interacting Bose and Fermi gases in harmonic potentials from modern textbooks \cite{Pethick2002,Pitaevskii2003,Pathria2011} and references therein, several important questions appear, which, in our opinion, should be addressed accordingly.
First, the widely used approach for studying such systems involves the semiclassical approximation.
However, it omits the lowest energy of the harmonic oscillator by setting it to zero and approximates the discrete spectrum by continuous distribution with a certain density of states.
By accepting this step, one neglects the corrections originating from delta-peak-like contributions of discrete energy levels and smooths out all of them assuming these are located close enough to each other to form a continuous structure.  
Second, the cold-atom experiments are restricted to the finite number of particles. 
Even though these numbers can be as high as several millions \cite{Dalfovo1999,Giorgini2008}, such restrictions can lead to 
observable corrections to thermodynamic quantities. 
Third, despite some estimates for Bose gases in anisotropic harmonic confinement and finite number of particles are present in the literature, the magnitude of these effects for Fermi gases remains unclear.

In the paper, we discuss the mentioned aspects by systematic analysis involving analytic (semiclassical) and exact numerical approaches for both quantum statistics.
We take ultracold lithium-6 and lithium-7 atomic gases in anisotropic harmonic trap as prototypical systems that allows to compare results and analyze the magnitude of the effects in corresponding experiments \cite{Truscott2001}.

\section{General equations and semiclassical approximation}\label{sec:2_semicl}
Let us briefly remind general quantum-mechanical results and main steps in the theoretical description of the system under study.
We describe an ideal atomic (bosonic or fermionic) gas of atoms with mass~$m$ in the external harmonic oscillator potential with the trapping frequencies $\omega_i$, $i=\{x,y,z\}$,
\begin{equation}
    U({\bf r}) = 
    \frac{m}{2}(\omega_x x^2+\omega_y y^2
    +\omega_z z^2).
\end{equation}
The eigenenergies of the Hamiltonian ${\cal H} = p^2/2m + U({\bf r})$ are given by
\begin{equation}
		\varepsilon_{n_{x}n_{y}n_{z}}
		=\sum_{i=\{x,y,z\}}
		\hbar\omega_{i}\left(n_{i}+\frac{1}{2}\right)
	\end{equation}
with $n_i=0,1,2,...$, while the eigenfunctions are
\begin{eqnarray}\label{eq:psi}
    	\psi_{n_{x}n_{y}n_{z}}(x,y,z)
    	&=&
    	\frac{1}{\pi^{3/4}}\prod\limits_{j=x,y,z}\frac{1}{(2^{n_j}\lambda_j n_{j}!)^{1/2}}
    	\\
    	\nonumber
    	&&\times
    	H_{n_{j}}\left(\frac{j}{\lambda_j}\right)\exp\left(-\frac{j^2}{2\lambda_j^2}\right),
\end{eqnarray}
where $\lambda_j=\left({\hbar}/{m\omega_{j}}\right)^{1/2}$ and $H_n (x)$ is the Hermite polynomial of the order $n$.
Below, we focus on the most common case of anisotropic harmonic trap with two characteristic frequencies: longitudinal $\omega_z$ and transverse $\omega_\perp\equiv\omega_{x,y}$.
For the purpose of the subsequent analysis, let us specify the ground-state wave function (for simplicity, we denote the ground state by the single index~0), which, according to Eq.~\eqref{eq:psi}, has the following form:
\begin{equation}\label{psi_0}
			\psi_{0}(\mathbf{r})
			=\left(\frac{m\bar{\omega}_{\rm tr}}{\pi\hbar}\right)^{3/4}\exp\left[-\frac{m}{2\hbar}
			(\omega_{\perp}
			(x^2+y^2)
			+\omega_{z}z^2)\right],
		\end{equation}
where 
$\bar{\omega}_{\rm tr}=(\omega_x\omega_y\omega_z)^{1/3}$ is the geometric mean frequency.

The distribution functions over the states with energies $\varepsilon_i\equiv\varepsilon_{n_{x}n_{y}n_{z}}$ are
\begin{equation}\label{eq:dist_func}
    f(\varepsilon_i) = \{\exp[(\varepsilon_i-\mu)/T]\pm1\}^{-1},
\end{equation}
where the signs ``$+$'' and ``$-$'' correspond to the Fermi and Bose statistics, respectively. Here and below, the temperature~$T$ is given in units $k_B=1$, unless specified otherwise.
In terms of the gas fugacity, 
\begin{equation}\label{eq:fugacity}
    z=\exp[\mu(T)/T],
\end{equation}
the distribution function~\eqref{eq:dist_func} can also be rewritten as $f(\varepsilon_i) = [z^{-1}\exp(\varepsilon_i/T)\pm1]^{-1}$.
The chemical potential~$\mu$ is determined from the normalization condition for the total number of atoms,
 \begin{equation}\label{eq:totN}
     N=\sum_{i=0}^{\infty}f(\varepsilon_i).
 \end{equation}

Let us also recall here general formulas for the main thermodynamic quantities.
The internal energy of a gas is given by
 \begin{equation}\label{eq:totE}
     E=\sum_{i=0}^{\infty}\varepsilon_i f(\varepsilon_i),
 \end{equation}
while the grand potential can be obtained from
 \begin{equation}\label{eq:Omega}
     \Omega=\mp T\sum_{i=0}^{\infty}\ln[1\mp f(\varepsilon_i)],
 \end{equation}
where the upper and lower signs correspond to fermions and bosons, respectively.
One can also relate the grand potential~$\Omega$ with the gas pressure~$P$ and the volume~$V$ as $\Omega=-PV$. However, in case of the harmonically-trapped gases, the volume $V$ becomes somehow ill-defined quantity, since the system does not have rigid spatial boundaries. There are ways to introduce effective spatial characteristics by relating them to trapping frequencies \cite{Pitaevskii2003,Romero2005}, but we omit this issue.

The specific heat and the entropy are usually calculated from the above formulas in a straightforward manner. In particular, the specific heat is determined as $C_N = (\partial E/\partial T)_N$.
The entropy is defined from the relation
\begin{equation}\label{eq:S}
    S = \frac{1}{T}(E-\Omega-\mu N).
\end{equation}
Alternatively, the entropy can be obtained by employing the derivatives $S=-(\partial \Omega/\partial T)_N-N(\partial \mu/\partial T)_N$, which is typically useful for verification purposes.

\subsection{Smoothed density of states}
The sums over the quantum states in all thermodynamic quantities are computed below numerically to perform exact analysis of quantum gases in the asymmetric harmonic potential. However, to compare exact results with analytical predictions, let us also introduce the semi-classical approximation \cite{Pathria2011,Pitaevskii2003,Pethick2002}, which allows one to replace the sums over the quantum numbers $n_x,n_y,n_z$ by corresponding integrals.
The key ingredient of this approximation is the smoothed density of states defined by 
\begin{equation}\label{eq:rho_e}
	\rho(\varepsilon)\approx\iiint_{0}^{\infty}
		\delta(\varepsilon-	\varepsilon_{n_{x}n_{y}n_{z}}+\varepsilon_0)dn_{x}dn_{y}dn_{z}.
\end{equation}
Performing successive integration and employing the well-known relation between the Dirac delta-function and the Heaviside step function, one obtains from Eq.~\eqref{eq:rho_e}
\begin{equation} \label{eq:ds_harm}
	\rho(\varepsilon)=\frac{1}{2}\varepsilon^2/(\hbar\bar{\omega}_{\rm tr})^3.
\end{equation}
Replacing the sum by an integral is usually a good choice if the density of
states is high, i.e., the energy spacing between nearest levels is small in comparison with temperature, $\hbar\omega_j\ll T$. 
However, in the regime $\hbar\omega_j\gtrsim T$
this may lead to notable inaccuracies.
It is also worth mentioning that the density of states in the form~\eqref{eq:ds_harm} does not account for the effects related to the zero-point energy $\varepsilon_0$.

In terms of $\rho(\varepsilon)$, the total particle number \eqref{eq:totN} can be written as
\begin{equation}\label{eq:totN_sc}
    N=\int_{0}^{\infty}f(\varepsilon)\rho(\varepsilon)d\varepsilon.
\end{equation}
This allows us to determine characteristic energy scales for quantum gases obeying both statistics.
In particular, for the Bose gas with the given $N$, one can specify the critical temperature~$T_{\rm c}$ below which the lowest single-particle state becomes macroscopically occupied, i.e., $N_0\neq0$ at $T\leq T_{\rm c}$. However, as we discuss below in more detail, this criterion is rigor only in the thermodynamic limit $N\to\infty$, where the semiclassical approach is valid.
For the trapped Fermi gas, the relevant thermodynamic quantity, the Fermi energy $\varepsilon_{\rm F}$, can be obtained from the energy of the highest occupied state in the limit $T\to0$. It can be determined for the given trap curvature and the total number of particles both exactly and within the semiclassical treatment.

\subsection{Approximation for the Bose gas}
According to the semiclassical approximation, see Eq.~\eqref{eq:ds_harm}, the lowest state has zero energy, $\min(\varepsilon)=0$. The chemical potential of the Bose gas cannot exceed this minimal value (otherwise, the distribution function for certain states becomes negative) and, thus, should be taken zero below $T_{\rm c}^{\rm sc}$. 
Evaluation of the integral~\eqref{eq:totN_sc} with the given density of states~\eqref{eq:ds_harm} and the Bose distribution function~\eqref{eq:dist_func} under condition $\mu(T= T_{\rm c}^{\rm sc})=0$ results in
\begin{equation}\label{eq:N_sc}
    N = (T_{\rm c}^{\rm sc}/ \hbar\bar{\omega}_{\rm tr})^3 \zeta(3),
\end{equation}
where $\zeta(s)$ is the Riemann zeta function.
This provides with the definition of the critical temperature
\begin{equation}\label{eq:Tc_sc}
    T_{\rm c}^{\rm sc}
    \approx0.94\hbar\bar{\omega}_{\rm tr}N^{1/3}.
\end{equation}
Here, $T_{\rm c}^{\rm sc}$ determines the temperature below which the lowest single-particle state ($\varepsilon=0$) becomes macroscopically occupied.

Within the semiclassical approximation, the chemical potential above the critical temperature can be determined by employing the equation for the total number of particles in the integral form \eqref{eq:totN_sc}. 
It can be written as
\begin{equation}
    \frac{1}{\zeta(3)}(T/T_{\rm c}^{\rm sc})^3{g_3(z)}=1,
\end{equation}
where $z$ is the gas fugacity~\eqref{eq:fugacity} and $g_{s}(z)=\Gamma^{-1}(s)\int_0^\infty dx x^{s-1}(z^{-1}e^x-1)^{-1}$ is the Bose-Einstein function
\cite{Pitaevskii2003,Pathria2011}.
The chemical potential, thus, can be determined on the whole temperature range by means of numerical root-search algorithms, as in Ref.~\cite{Sotnikov2017LTP} for homogeneous gases.

Below $T_{\rm c}^{\rm sc}$ and at $\mu=0$, the right-hand side of Eq.~\eqref{eq:totN_sc} also determines the number of particles in the thermal component $N_T$. Therefore, replacing $N$ by $N_T$ there and using that $N=N_0+N_T$, we obtain the number of particles, which occupy the lowest single-particle state at $T\leq T_{\rm c}$,
\begin{equation}\label{eq:N0}
    N_0=\left[1-(T/T_{\rm c}^{\rm sc})^3
    \right]N.
\end{equation}

The gas density can be expressed as a sum of the condensate and the thermal components,
\begin{eqnarray}\label{eq:dens_bos}
   n(\mathbf{r})
    = N_0|\psi_0({\bf r})|^2 + n_{T}(\mathbf{r}),
\end{eqnarray}    
where $\psi_0$ is the ground-state wave function~\eqref{psi_0} and $N_{0}$ is determined by Eq.~\eqref{eq:N0}.
To obtain the explicit form of the second term, one needs to 
replace the quantized energies by the classical expression $\varepsilon({\bf p},\mathbf{r})={\bf p}^2/(2m)+U(\mathbf{r})$ in the distribution function \eqref{eq:dist_func}. The subsequent integration of $n_T({\bf r})=(2\pi\hbar)^{-3}\int d^3{p} f({\bf p},{\bf r})$ over momentum ${\bf p}$ yields
\begin{eqnarray}\label{eq:dens_bosT}
	n_{T}(\mathbf{r})
		=\lambda_T^{-3}
		g_{3/2}\left(\exp{\left[\frac{\mu-U(\mathbf{r})}{T}\right]}\right),
\end{eqnarray}
where $\lambda_T=\sqrt{{2\pi\hbar^2}/{mT}}$ is the thermal de Broglie wavelength.

Within the semiclassical approach, the thermodynamic quantities can be obtained in terms of the Bose-Einstein functions 
by replacing the sums by the integrals with the given density of states.
In particular, according to Eqs.~\eqref{eq:totE} and \eqref{eq:ds_harm}, the total energy reads
 \begin{equation}\label{eq:totE_sc}
     E=3T\left(\frac{T}{\hbar\bar{\omega}_{\rm tr}}\right)^{3}g_4(z).
 \end{equation}
The grand potential can be obtained from Eqs.~\eqref{eq:Omega} and \eqref{eq:ds_harm},
 \begin{equation}\label{eq:Omega_sc}
     \Omega=-T\left(\frac{T}{\hbar\bar{\omega}_{\rm tr}}\right)^{3}
     g_4(z).
 \end{equation}
 Taking derivative $\left(\partial E/\partial T\right)_N$ of the internal energy \eqref{eq:totE_sc} and using recurrent relation $\partial g_s(z)/\partial z=g_{s-1}(z)/z$ \cite{Pathria2011}, we obtain the specific heat,
\begin{eqnarray}\label{eq:C_sc_bos}
        \frac{C_N}{N}=
        \begin{cases}
            \frac{12\zeta(4)}{\zeta(3)}
            \left(\frac{T}{T_c^{\rm sc}}\right)^3,~
            T\leq T_c^{\rm sc}; 
        \\
            \frac{1}{\zeta(3)}
        \left(\frac{T}{T_c^{\rm sc}}\right)^3
        \left(12g_4(z)-\frac{9g_3^2(z)}{g_2(z)}\right),
        ~T>T_c^{\rm sc}.
        \end{cases}
\end{eqnarray}
The entropy can be expressed in the following form:

\begin{equation}\label{eq:S_sc_bos}
    \frac{S}{N}=\frac{4}{\zeta(3)}\left(\frac{T}{T_{\rm c}^{\rm sc}}\right)^3g_4(z)-\ln{z}.
\end{equation}
From Eqs.\eqref{eq:C_sc_bos} and \eqref{eq:S_sc_bos} we can ensure that the specific heat exhibits the discontinuity $\Delta C_N = -9N\zeta(3)/\zeta(2)$ at $T=T_{\rm c}^{\rm sc}$ ($z=1$), while the entropy remains a continuous function. 

\subsection{Approximation for the Fermi gas}
For the Fermi gas at $T=0$ the mean occupation number of the single-particle state is equal to 
\begin{equation}\label{eq:n_step}
	f(\varepsilon)
	=\left\{
	\begin{aligned}
		1, &\quad\varepsilon<\mu(0); \\ 
		0, &\quad\varepsilon>\mu(0). \\
	\end{aligned}
	\right.
\end{equation}
 Taking into account \eqref{eq:n_step}, we see that at $\varepsilon<\mu(0)$ all energy states are occupied according to the Pauli exclusion principle. 
 The highest occupied energy state refers to  the Fermi energy $\varepsilon_{\rm F}=\mu(0)$.

 Integrating the density $\rho(\varepsilon)$ of single-particle states in the framework of semiclassical approximation, see Eq.~\eqref{eq:ds_harm},  we obtain the equation defining the Fermi energy~$\varepsilon_{\rm F}^{\rm sc}$ (or, equivalently, the Fermi temperature~$T_{\rm F}^{\rm sc}$),
\begin{equation}\label{eq:energy_F}
    \varepsilon_{\rm F}^{\rm sc}=\left(6N\right)^\frac{1}{3}\hbar \bar{\omega}_{\rm tr}.
\end{equation}
Within the same approach, the chemical potential on the whole temperature range can be determined by employing the equation for the total number of particles in the integral form \eqref{eq:totN_sc}, 
\begin{equation}
    6\left(T/T_{\rm F}^{\rm sc}\right)^3f_3(z)=1,
\end{equation}
where $z$ is the gas fugacity~\eqref{eq:fugacity} and
$f_{s}(z)=\Gamma^{-1}(s)\int_0^\infty dx x^{s-1}(z^{-1}e^x+1)^{-1}$ is the Fermi-Dirac function.

The density distribution of the Fermi gas in a harmonic trap can be calculated similarly to the density of the Bose gas, see the text above Eq.~\eqref{eq:dens_bosT},
\begin{equation}\label{eq:dens_fer}
    n(\mathbf{r})
		=\lambda_T^{-3}
		f_{3/2}\left(\exp{\left[\frac{\mu-U(\mathbf{r})}{T}\right]}\right).
\end{equation}

Furthermore, we can obtain equations, which describe the main thermodynamic characteristics of the Fermi gas. 
The internal energy~\eqref{eq:totE} reads
 \begin{equation}\label{eq:totE_sc_ferm}
     E=3T\left(\frac{T}{\hbar\bar{\omega}_{\rm tr}}\right)^{3}f_4(z).
 \end{equation}
The grand potential~\eqref{eq:Omega} is 
\begin{equation}\label{eq:pot_sc_ferm}
    \Omega=-T\left(\frac{T}{\hbar\bar{\omega}_{\rm tr}}\right)^{3}f_4(z),
\end{equation}
which is connected to the total energy by the relation $E= -3\Omega$.

Contrary to the specific heat of the Bose gas, which has a discontinuity at $T=T_{\rm c}^{\rm sc}$, the one of the Fermi gas is a continuous function defined on the whole temperature range in the following way: 
\begin{equation}\label{eq:C_sc_fer}
     \frac{C_N}{N}= 6\left(\frac{T}{T_{\rm F}^{\rm sc}}\right)^3
        \left[12f_4(z)-\frac{9f_3^2(z)}{f_2(z)}\right].
\end{equation}
Using Eqs.~\eqref{eq:S}, \eqref{eq:totE_sc_ferm}, and \eqref{eq:pot_sc_ferm} we can obtain the equation for the entropy of the Fermi gas,
\begin{equation}\label{eq:S_sc_fer}
    \frac{S}{N}=24\left(\frac{T}{T_{\rm F}^{\rm sc}}\right)^3f_4(z)-\ln{z}.
\end{equation}

In the next sections, we employ the provided analytic expressions for thermodynamic quantities as a convenient visual reference for numerical dependencies obtained within exact techniques for systems with finite number of particles and quantized energy spectrum.

\section{Total particle number and chemical potential}\label{sec:3_tot_N}

Let us point out an important consequence of equivalence of the trapping frequencies ($\omega_x=\omega_y$) that allows to simplify the succeeding numerical analysis beyond the semiclassical approximation.
In particular, Eq.~\eqref{eq:totN} can be transformed to the following expression:
 \begin{equation}\label{eq:totN_quant}
     N=\sum_{q=0}^{\infty}\sum_{n_z=0}^{\infty}\frac{1+q}{
     \exp\{[\varepsilon_0+\hbar\omega_z(n_z+\chi q)-\mu]/T\}
     \pm1},
 \end{equation}
where we introduced the quantum number $q=n_x+n_y$. In the given form, the factor $(1+q)$ in the numerator corresponds to the degeneracy of the harmonic-oscillator states due to equal trapping frequencies along two spatial directions ($\omega_{x,y}=\omega_\perp$).
For convenience, we also denoted the ratio of the longitudinal and transverse frequencies by $\chi=\omega_{\perp}/\omega_z$.
Compared to the general relation~\eqref{eq:totN}, Eq.~\eqref{eq:totN_quant}  significantly reduces computational cost in the numerical analysis of thermodynamic quantities, thus, achieving sufficient accuracy for systems with $N\sim10^6$ atoms or higher in the temperature range corresponding to the quantum degeneracy regime.

Next, let us emphasize another important effect for the ideal Bose gas consisting of the finite number of particles.
In particular, from Eq.~\eqref{eq:totN_quant} we see that, as soon as the total number $N$ of particles in the system is taken finite (and $T>0$),
the chemical potential {\it cannot exceed or even become equal} to the minimal energy $\varepsilon_0$, $\mu(T)<\varepsilon_0$.
In other words, for any finite $N$ one can always determine the chemical potential, such that
\begin{equation}\label{eq:delta_def}
    \mu(T)=\varepsilon_0 - \delta_N(T),
\end{equation}
where $\delta_N(T)\geq0$, whereas the equality holds only in two cases: (i) at $T=0$ or (ii) at $N=\infty$ and $T\leq T_{\rm c}^{\rm sc}$.

The asymptotic behavior of the chemical potential with $\delta_N>0$ corresponds to the fact that for the system with a finite number of particles \emph{there is no conventional phase transition} associated with the discontinuities of thermodynamic quantities at the critical point. 
The ``exact'' critical temperature~$T_c$ cannot be determined in the mathematically strict manner, thus, we avoid this notation below. 

Let us estimate the finite-number correction $\delta_N(T)$ in the low-temperature regime, i.e., at $T\ll T_c^{\rm sc}$. Obviously, by taking the first term in Eq.~\eqref{eq:totN_quant} ($q=n_z=0$), i.e., $N_0\approx N$, with $\mu(T)$ given by Eq.~\eqref{eq:delta_def} we obtain
\begin{equation}
    \exp[\delta_N(T)/T]-1 = N^{-1}.
\end{equation}
Next, assuming that the strong inequality $N\gg1$ remains valid, the Taylor expansion yields
\begin{equation}\label{eq:delta_lin}
    \delta_N(T) \approx T/N.
\end{equation}

\begin{figure}
\includegraphics[width=\linewidth]{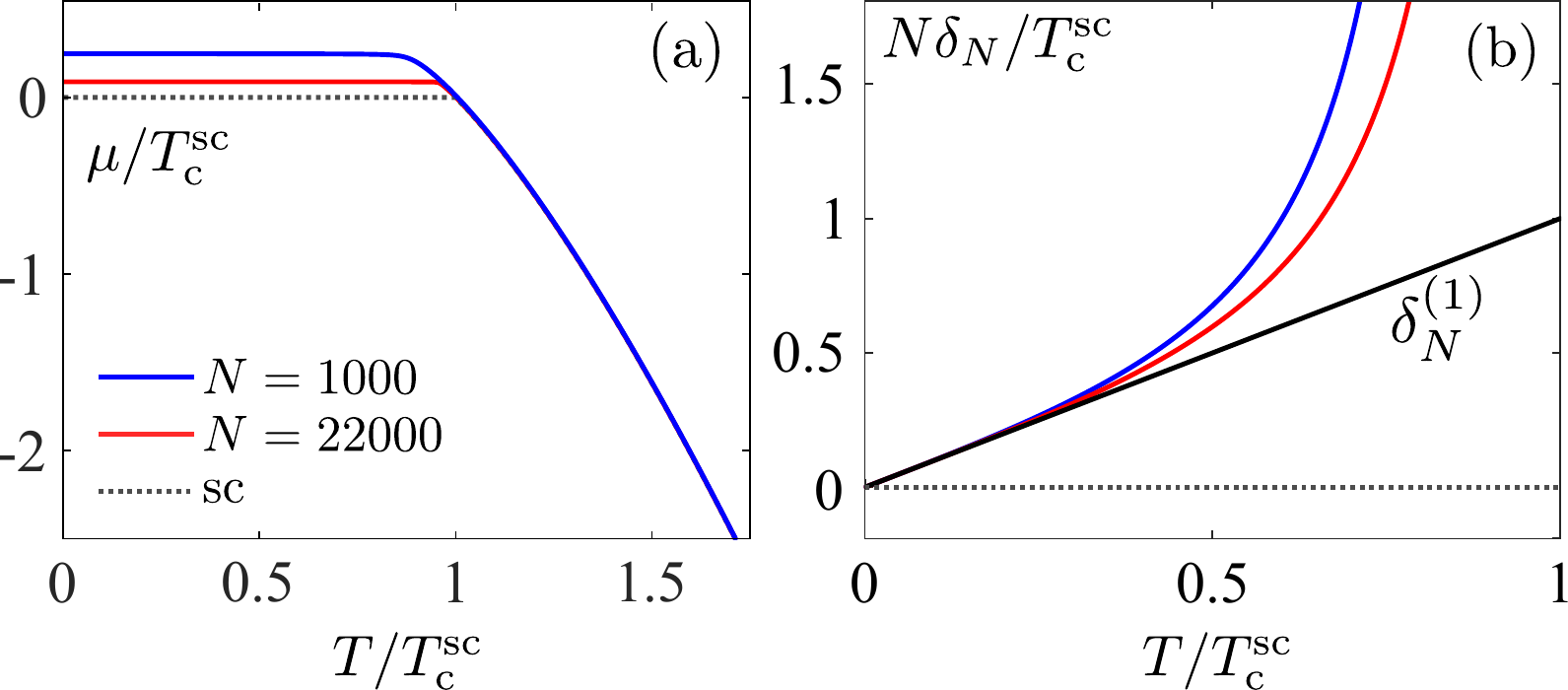}
    \caption{The chemical potential (a) and its finite-particle correction (b) of the trapped Bose gas as functions of temperature. 
    The colored solid lines correspond to numerical results, while the dotted line is obtained within the semiclassical approach. $\delta_N^{(1)}$ is the linear correction~\eqref{eq:delta_lin}. }
    \label{fig:mu_b}
\end{figure}

In Fig.~\ref{fig:mu_b}(a) we plot $\mu(T)$ dependencies for a system consisting of finite number of particles and compare these to the semiclassical results.
The saturated nonzero values of the chemical potentials at a given value of $N$ and $T<T_{\rm c}^{\rm sc}$ correspond (up to the  correction $\delta_N$) to the zero-point energy~$\varepsilon_0$. Let us emphasize that in the saturated regime $\mu(T)$ are not constant but slowly decreasing functions of temperature. In contrast, the semiclassical approximation yields $\varepsilon_0=0$.
The correction $\delta_{N}(T)$ is shown separately in Fig.~\ref{fig:mu_b}(b) and agrees well with the low-$T$ expansion~\eqref{eq:delta_lin} in the limit $T\to0$.  

Note that with an increase of $T$ exact results departure from the linear dependence~\eqref{eq:delta_lin}.
Among the possible reasons, we verified that the second order Taylor expansion does not lead to a better match of $\delta^{(2)}_N$ with the curves at finite $N$. 
The true reason is that the further corrections come from the next few terms in Eq.~\eqref{eq:totN_quant} corresponding to the excited states, as we observe by means of exact numerical analysis. The complete match is achieved if all occupied levels are taken into account.

\begin{figure}
\includegraphics[width=\linewidth]{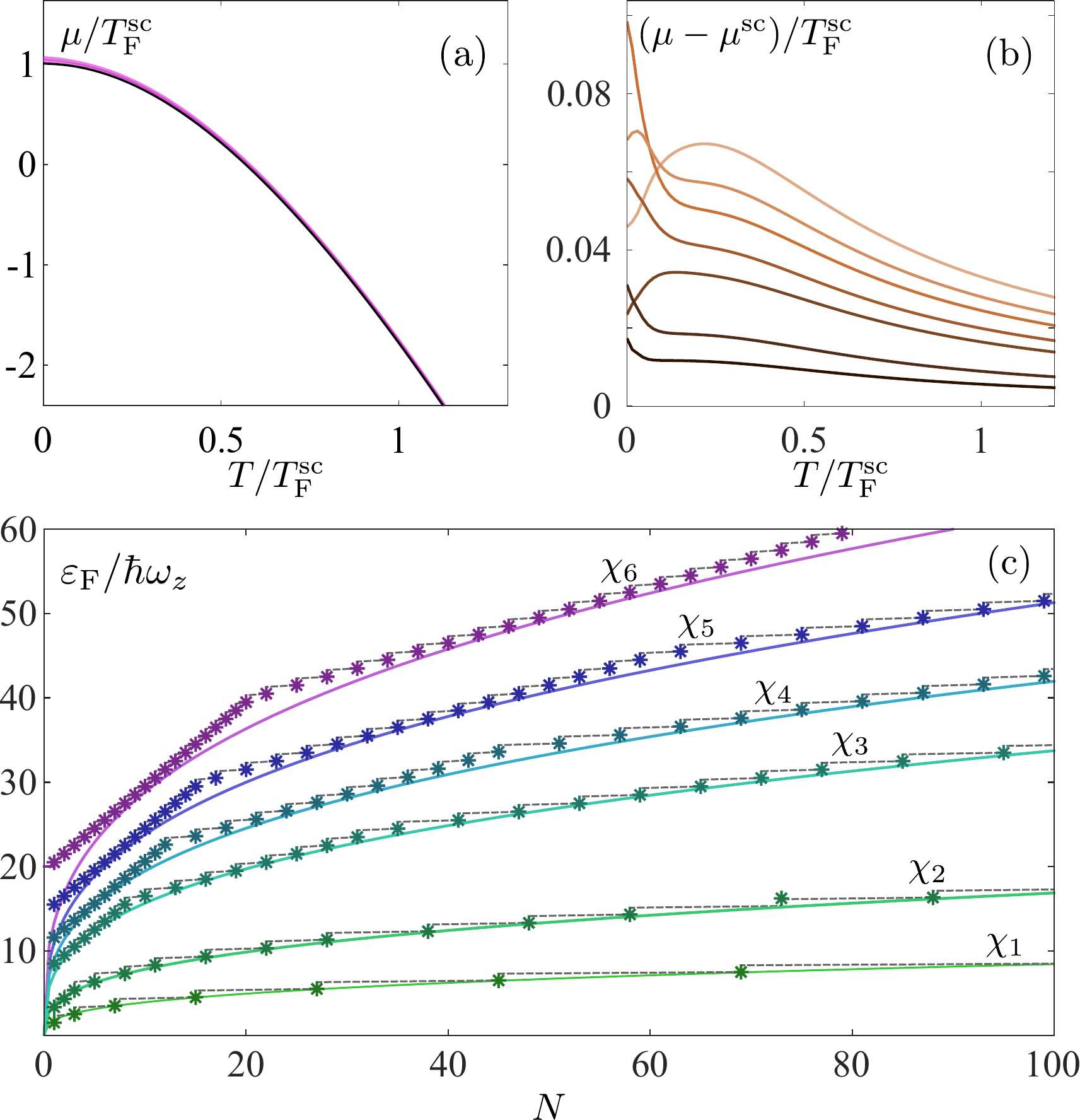}
    \caption{Temperature dependencies of the chemical potential (a) and its relative difference with respect to semiclassical results (b) of the Fermi gas. In panel (b), the lines correspond to $N=7,9,11,15,20,50,100$ from top to bottom on the right-hand side at $\chi=11.1$.
    Dependencies of the Fermi energy on the total particle number for different trap anisotropies (c): $\{\chi_1,...,\chi_6\}=\{1,\sqrt{8},8,11.1,15,20\}$. 
    Solid lines correspond to the semiclassical approach, while dots are numerical results connected by dashed lines for sake of visibility.}
    \label{fig:mu_f}
\end{figure}
Accounting for the finite number of particles in a Fermi gas leads to relatively small corrections to the chemical potential. 
In particular, for $N<100$, these corrections become only visible at $T\to 0$ due to observable corrections to the Fermi energy originating from the discrete energy structure, see Fig.~\ref{fig:mu_f}(b,c).

In Fig.~\ref{fig:mu_f}(a), the slightly broadened pink curve {summarizes the temperature dependencies} of the chemical potential for a small number of particles ($N<100$), while the black line corresponds to the case $N\gtrsim1000$, which we identify with the large-$N$ limit, or, equivalently, the limit of validity of the semiclassical approximation. 
The disagreement between numerical and semiclassical values of the Fermi energy are noticeable only at $N<100$, see Fig.~\ref{fig:mu_f}(c), where trap anisotropies $\chi_{2},\chi_{4}$ are taken from Refs.~\cite{Truscott2001,Ensher1996}. With a further increase in $N$ the difference vanishes.
Therefore, below, while describing the properties of the Fermi gas with $N\gtrsim1000$,  we omit the index ``sc'' for brevity.

The obtained explicit dependencies of the chemical potentials on the temperature are crucial and allow one further to construct all relevant thermodynamic characteristics as functions of temperature similar to homogeneous gases~\cite{Sotnikov2017LTP}. Since these are now calculated both within the semiclassical treatment and exactly for the finite number of particles in a trap, the differences in the behavior should also be noticed in other observables.

\section{Spatial density distributions}\label{sec:4_distr}

\begin{figure}
\includegraphics[width=\linewidth]{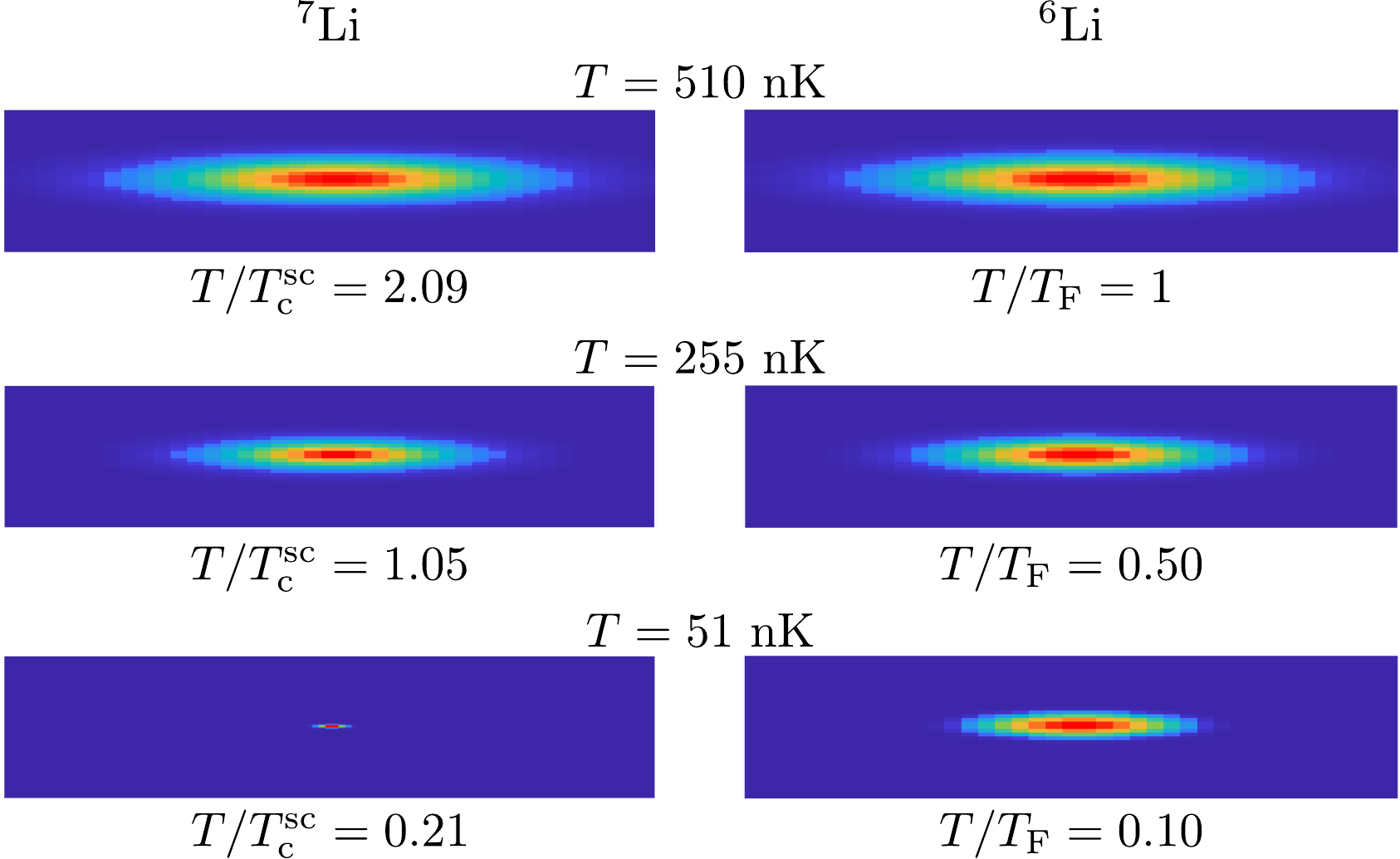}
    \caption{Color contour plots characterizing the density distributions of atomic Bose (left) and Fermi (right) gases.
    The trapping frequencies $\omega_i$ are taken the same as in Ref.~\cite{Truscott2001} and the total number of atoms is fixed to $N=2.2\times 10^4$, as in one of experimental realizations.
    The size of each image is 0.12$\times$0.7~mm along the vertical and horizontal direction, respectively.}
    \label{fig:3_color_dens}
\end{figure}

\begin{figure}
\includegraphics[width=\linewidth]{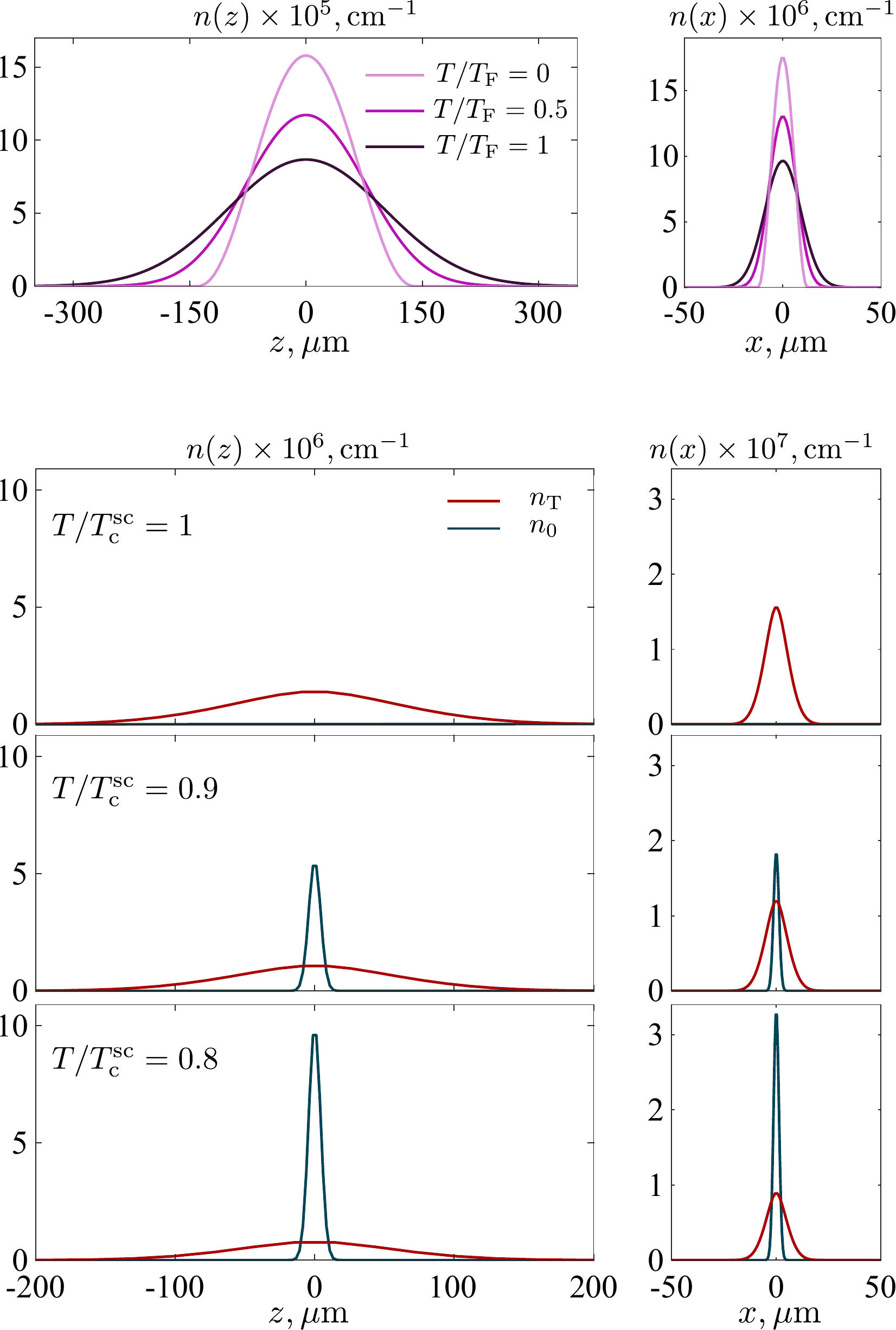}
    \caption{Density distributions of the Fermi (upper row) and the Bose (three lower rows) gases in the trap for different temperature values. 
    For the Bose gas, the thermal and the condensate density components are illustrated separately. 
    The total number of particles is kept fixed everywhere, $N=22000$.
   }
    \label{fig:nx_nz}
\end{figure}

The bunching and antibunching effects for Bose and Fermi gases, respectively, were nicely demonstrated on the example of atomic $^7$Li and $^6$Li gases in the experiment~\cite{Truscott2001}.
Due to natural experimental limitations in the original work, it was difficult to keep the number of atoms fixed and to cool the Bose gas significantly below the critical temperature.
For the given experimental parameters, we perform our theoretical analysis with the fixed number of particles in the whole temperature range and show our results in Fig.~\ref{fig:3_color_dens}.
The color-coded images are obtained by the integration of the corresponding density distributions \eqref{eq:dens_bos} and \eqref{eq:dens_fer} along one of transverse directions 
\footnote{In Fig.~\ref{fig:3_color_dens} the colormaps are given in arbitrary units; for quantitative analysis see Fig.~\ref{fig:nx_nz}}.
For the Bose gas at temperature slightly above $T_{\rm c}^{\rm sc}$, we find a good agreement for spatial extents of the atomic cloud with the experiment~\cite{Truscott2001}, where the data was given for the same number of particles ($N=2.2\times 10^4$) and trap curvature.
 
At temperatures exceeding $T_{\rm F}$ (or approximately twice $T_{\rm c}^{\rm sc}$), the density distributions of quantum gases in harmonic trap become almost indistinguishable one from another (see upper row of Fig.~\ref{fig:3_color_dens}) and can be approximated by classical Boltzmann statistics.
As we show in the next section, a similar behavior in the high-$T$ limit holds for other relevant characteristics of quantum gases.

In Fig.~\ref{fig:nx_nz}, we also provide quantitative dependencies of the column density in different spatial directions demonstrating the condensate peak and melting of the Fermi surface, in accordance with Eqs.~\eqref{eq:dens_bos}, \eqref{eq:dens_bosT}, and \eqref{eq:dens_fer}, respectively.
In particular, the fixed value of the particle number~$N$ leads to the same areas under the curves for all cases. This means that with the temperature  decrease the Fermi surface becomes more rigid and approaches a certain saturated value at $T=0$ due to the Fermi pressure, whereas in the Bose gas a characteristic narrow condensate peak starts to form.

Note that for the Fermi gas in the limit $T\ll T_{\rm F}$, the Sommerfeld-expansion approximation can be applied to construct the distribution of the gas density.
 We obtain, in particular,
 \begin{equation}
	n(\mathbf{r})\approx\frac{\sqrt{2}qm^{3/2}}{3\pi^2\hbar^3}
	\Delta_{\bf r}^{3/2}+\frac{qm^{3/2}T^2}{12\sqrt{2}\hbar^3\Delta_{\bf r}^{1/2}} + {\cal O}(T^4), 
\end{equation}
where $\Delta_{\bf r}\equiv(\mu-U(\mathbf{r}))$. 
It is worth mentioning that this approximation requires $\Delta_{\bf r}>0$ and fails faster at the edges than in the trap center, since $\Delta_{\bf r}$ becomes of the order of $T$ faster in this region with the temperature increase.

In case of the Bose gas, the density distributions are shown in lower panels of Fig.~\ref{fig:nx_nz}. At $T=T_{\rm c}^{\rm sc}$ almost all particles occupy the excited states, which corresponds to the horizontal line with $n_0\approx0$~\footnote{Exact numerical calculation yields $N_0\approx11$ at $T=T_{\rm c}^{\rm sc}$ for the chosen set of trap parameters and $N=22000$.}. 
Given that at $T<T_{\rm c}^{\rm sc}$ bosons start to macroscopically occupy the lowest single-particle state, the condensate peak develops. 
We observe a rapid peak growth with a small decrease of $T$. It is explained by the cubic temperature dependence of the number of particles in excited states, see Eq.~\eqref{eq:N0}, in contrast to the uniform gas with $N_0\propto[1-(T/T_c)^{3/2}]$.

\section{Thermodynamic characteristics at finite particle number}\label{sec:5_thermodyn}

\begin{figure}
\includegraphics[width=\linewidth]{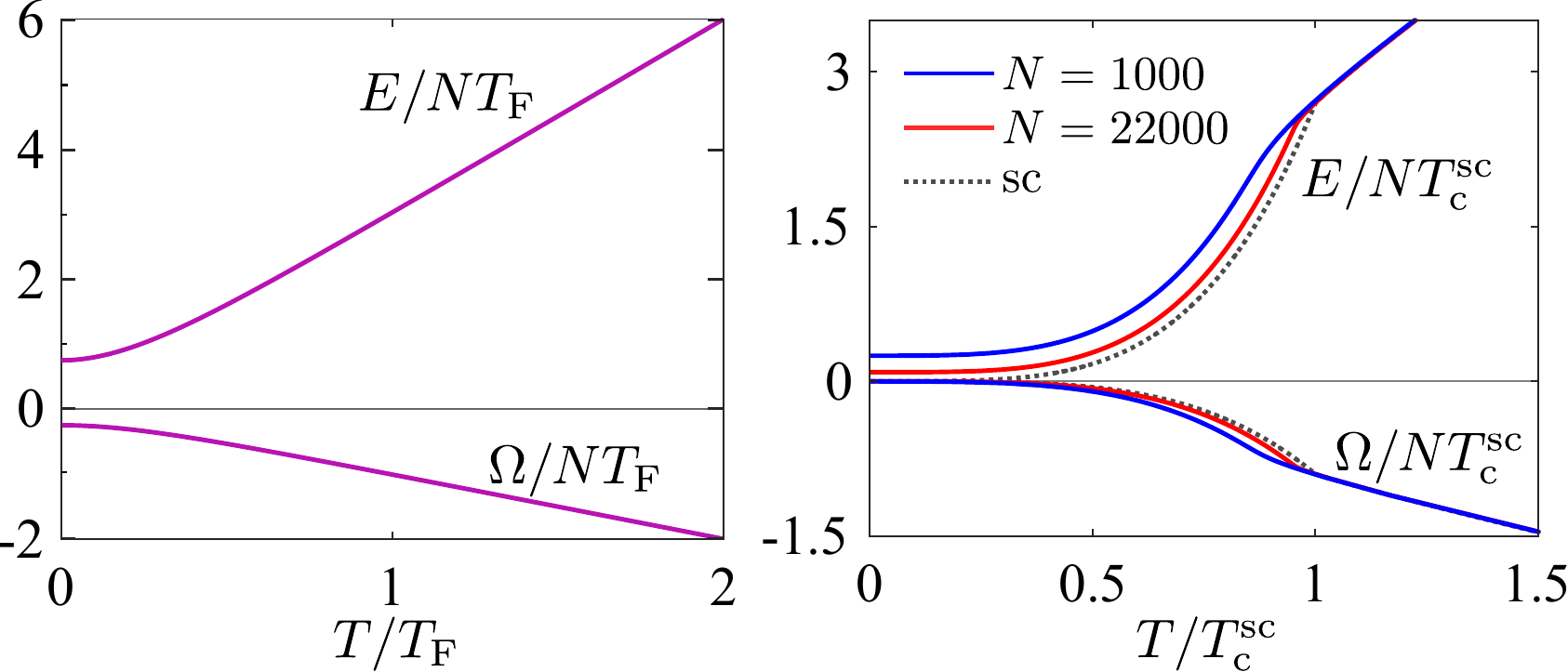}
    \caption{Temperature dependencies of the total energy $E$ and the grand potential $\Omega$ for the Fermi (left) and the Bose (right) gases. The solid lines represent numerical results, while the dotted line corresponds to the semiclassical approximation.}
    \label{fig:E_Omega}
\end{figure}

As we already mentioned, by following the semiclassical treatment, 
the energy and the grand potential are related by $ E~=-3\Omega$, see Eqs.~\eqref{eq:totE_sc} and \eqref{eq:Omega_sc}, as well as Eqs.~\eqref{eq:totE_sc_ferm} and \eqref{eq:pot_sc_ferm}. 
For the Fermi gas with the given $N=1000$ and $N=22000$, this relation holds with a good accuracy in the whole temperature range.
In particular, the total energy $E\to 3/4N\varepsilon_{\rm F}$ and the grand potential $\Omega\to -1/4N\varepsilon_{\rm F}$ as $T\to0$, see Fig.~\ref{fig:E_Omega}, where the corresponding lines overlap.
Here, the non-vanishing $\Omega$ is associated with the Fermi pressure in homogeneous gases.

However, in case of the Bose gas, the relation between $E$ and $\Omega$ is no longer valid according to the quantum approach, in particular, due to non-vanishing lowest-state energy $\varepsilon_0$. This effect can be seen for the Bose gas in Fig.~\ref{fig:E_Omega}, where $E\to N\varepsilon_0$, while $\Omega\to0$ in the zero-temperature limit. 
Furthermore, as soon as we go beyond the semiclassical approach ($N$ is finite), the energy becomes a smooth function of temperature, i.e., a characteristic kink in the Bose gas at $T\approx T_{\rm c}^{\rm sc}$ disappears.
The reason of such behavior is discussed below.

\begin{figure}
\includegraphics[width=\linewidth]{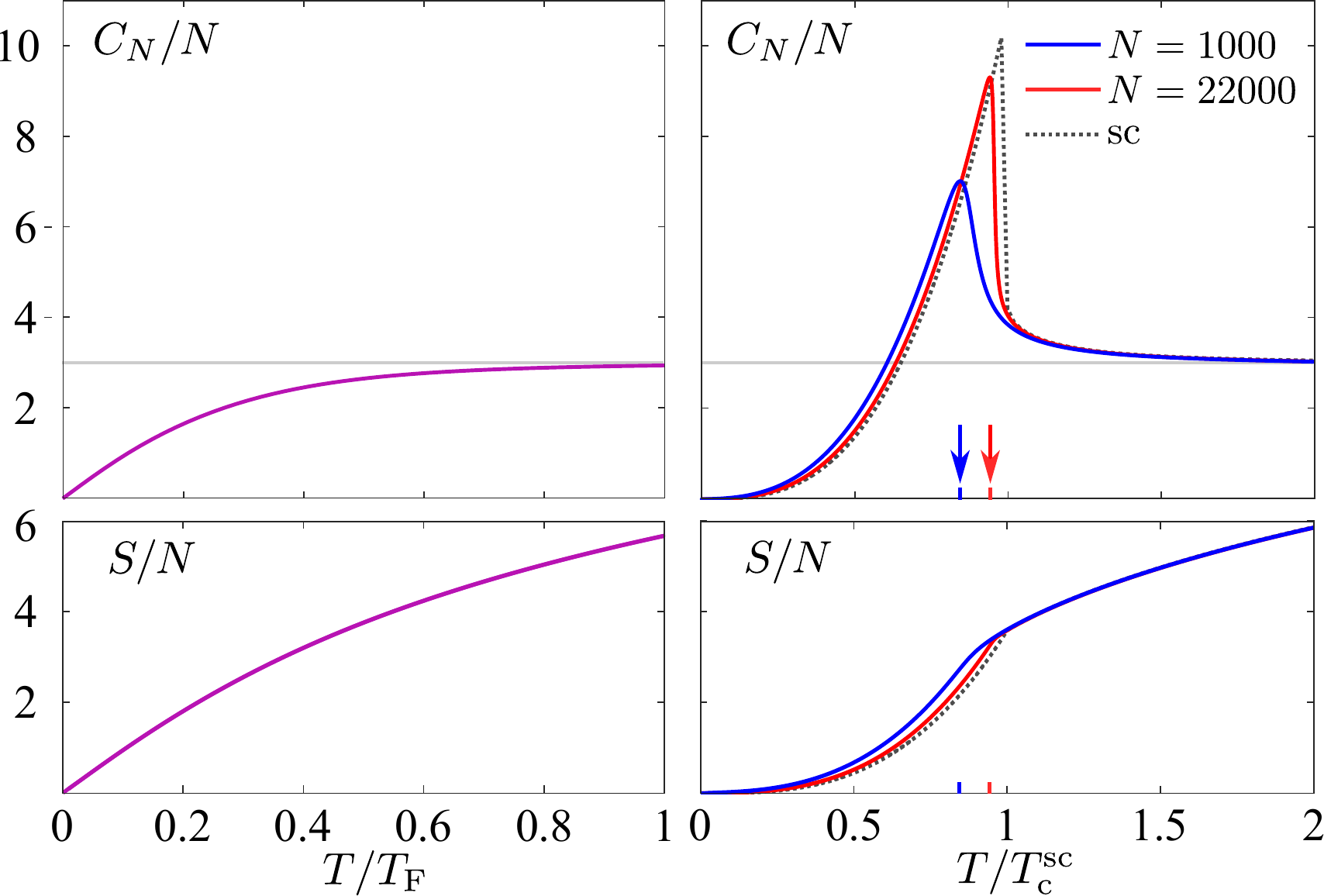}
    \caption{Temperature dependencies of the specific heat (upper row) and the entropy (lower row) for the Fermi (left) and the Bose (right) gases. The solid lines correspond to different number of particles while the dotted line corresponds to the semiclassical approximation. The vertical arrows indicate the positions of the $C_N$ maxima. }
    \label{fig:C_S}
\end{figure}

For the Fermi gas, both the specific heat and the entropy are continuous functions of temperature that agrees both with Eq.~\eqref{eq:S} and Eqs.~\eqref{eq:C_sc_fer} and \eqref{eq:S_sc_fer}; for the chosen $N$ the deviations between methods are negligible small, see Fig.~\ref{fig:C_S}, where the corresponding curves overlap.
At $T\approx T_{\rm F}$ the specific heat $C_N$ saturates and approaches $3NT$, in agreement with the equipartition theorem valid for classical limit.

As we also observe from Fig.~\ref{fig:C_S}, the temperature dependence of the entropy $S$ of the trapped Bose gas is qualitatively similar to the temperature dependence of the total energy $E$, see Eqs.~\eqref{eq:S} and \eqref{eq:S_sc_bos}, i.e., there is a characteristic kink at $T=T_{\rm c}^{\rm sc}$ only if one applies the semiclassical approach. At higher temperatures the numerical results demonstrate a good agreement with the semiclassical ones and reproduce the classical behavior $S^{\rm cl}/N=4+\ln[(T/\hbar\bar{\omega}_{\rm tr})^3/N]$ valid for both statistics.

In accordance with Eq.~\eqref{eq:C_sc_bos}, at $T=T_{\rm c}^{\rm sc}$ the specific heat of the Bose gas has a discontinuity, which indicates the first-order phase transition. However, numerical calculations give us qualitatively different temperature dependence of the specific heat, see Fig.~\ref{fig:C_S}. 
At finite $N$ this becomes a continuous function.
Note that the peak softening of the curve depends on the total particle number of the gas and implies the absence of the first-order phase transition. Therefore, the definition of the critical temperature is not valid anymore.
We suggest that the position of the maximum of $C_N$ can be put into correspondence with the transition temperature. The vertical arrows in Fig.~\ref{fig:C_S} demonstrate that the latter is shifted toward smaller values with the decrease of $N$.

Let us now discuss in more detail corrections to the critical temperature in the system under study.
Some steps to improve the semiclassical approximation 
consist of accounting for the zero-point energy \cite{Pethick2002}, or, alternatively, introducing the effective density of states for trapped bosons \cite{Grossmann1995}.
As for the first method, it yields the shift~${\Delta T}$ of the original critical temperature $T_{\rm c}^{\rm sc}$ to the lower values in accordance with the relation
\begin{equation}\label{eq:Pethick}
    \frac{\Delta T}{T_{\rm c}^{\rm sc}}\approx
    -0.73\frac{\omega_{\rm m}}{\bar{\omega}_{\rm tr}}N^{-1/3},
\end{equation}
where the arithmetic mean
    $\omega_{\rm m}=(\omega_x+\omega_y+\omega_z)/3$. 

\begin{figure}
\includegraphics[width=\linewidth]{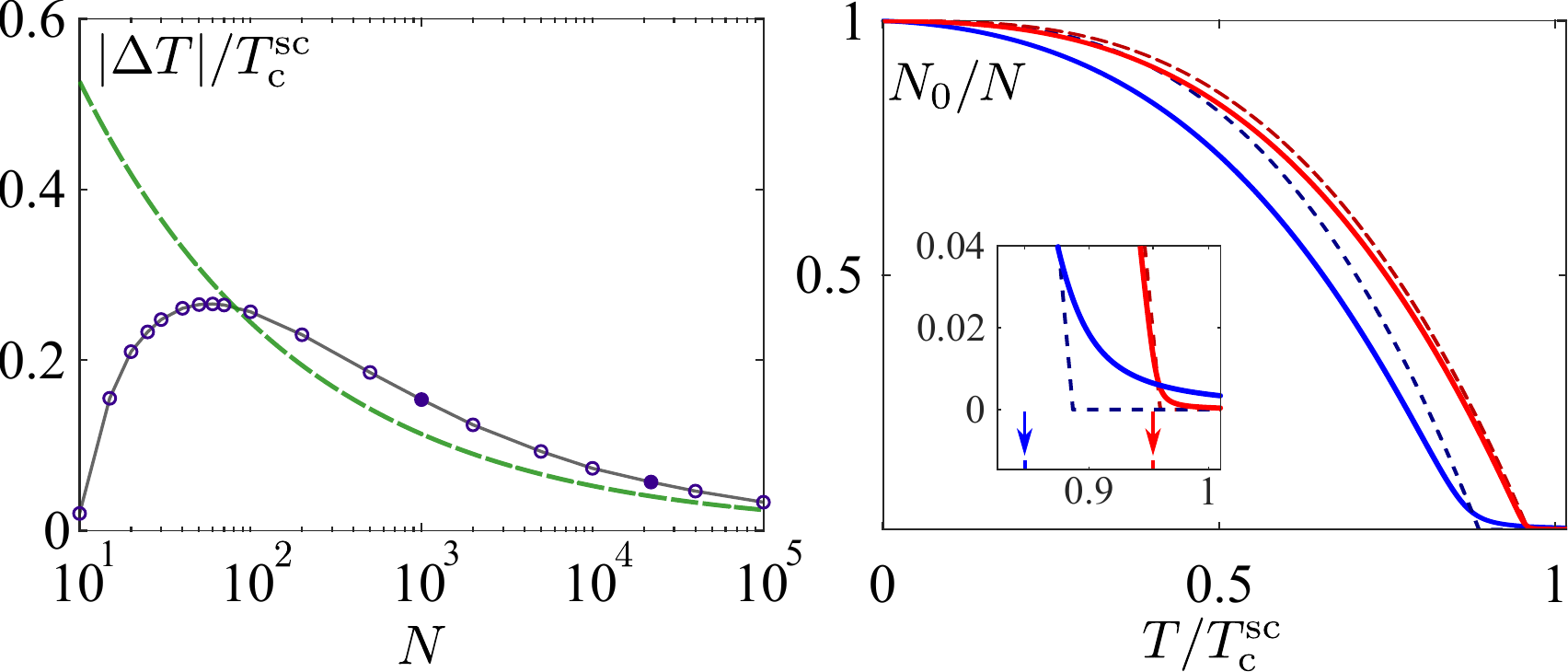}
    \caption{Left panel: corrections to the critical temperature of the Bose gas as a function of total particle number at $\chi=11.1$. 
    Right panel: temperature dependence of the normalized particle number in the lowest state with $\varepsilon_0$. The dashed lines correspond to the approximation \eqref{eq:Pethick}, while the solid lines and points are numerically obtained data (see main text).}
    \label{fig:Tc_corr}
\end{figure}

In Fig.~\ref{fig:Tc_corr} we plot the estimated correction given by Eq.~\eqref{eq:Pethick}, as well as the calculated corrections from the exact numerical analysis of the specific heat discussed above.
We observe that the widely-used approximation~\eqref{eq:Pethick},
even with further corrections \cite{Jaouadi2011},
systematically underestimates the effective transition temperature at $N\gtrsim100$.
The estimated critical temperature tends to the exact results only at $N>10^5$ (for the regime of small $N$, see \footnote{At $N<100$ the temperature dependencies of the specific heat cease to have a vivid maximum associated with the changes in the many-body state. 
}).
Note that in the experimental studies, e.g., in Ref.~\cite{Ensher1996,Griesmaier2005}, one can also notice the deviations between the experimentally measured data and the theoretical predictions relying on Eq.~\eqref{eq:Pethick}.
To show the effect more explicitly, in Fig.~\ref{fig:Tc_corr} we also plot dependencies of the number of particles in the ground state $N_0/N$.
Note that in addition to the physically justified bend of the curves (indicating absence of the phase transition), we also observe smaller values of $N_0/N$ in the intermediate range of temperatures due to corrections from the discrete structure of the first excited levels.
Similar behavior of $N_0/N$ was also pointed out in theoretical studies of ideal gas with finite $N$ in isotropic 3D harmonic potential \cite{Ketterle1996}.

\section{Conclusion}\label{sec:conclusion}

We studied equilibrium properties of the harmonically trapped ideal Bose and Fermi gases in the quantum degeneracy regime.
The analysis of thermodynamic characteristics of gases was performed  by means of the semiclassical approach and compared with exact numerical results for a finite number of particles. 

We examined the limits of applicability of the semiclassical approach widely employed in the literature.
To this end, we constructed exact temperature dependencies of the chemical potentials in systems consisting of finite number of trapped atoms.
For a Fermi gas, we demonstrated deviations in the Fermi energy values originating from a discrete level structure and showed that these appear only for a small number of particles. For a Bose gas, we observed characteristic softening of phase transition features, which contrasts to the semiclassical predictions \cite{Pethick2002,Pathria2011,Pitaevskii2003} and related approximations \cite{Jaouadi2011,Yukalov2005}.
We provided a more accurate methodology of determining corrections to the critical temperature due to finite number of atoms.
At the same time, we point out that the concept of phase transition in these systems is not valid in a strict sense due to smooth character of all thermodynamic functions.

Our results are valuable from the point of view of theoretical approaches and experiments aiming to accurately determine shifts in the transition temperature in weakly-interacting Bose gases.
The origin of these shifts is typically twofold: the first one comes from the finite number of particles and another one is associated with the interaction effects \cite{Giorgini1996}. 
By improving the description of systems with finite number of particles, one can give more accurate predictions on the impact of interaction effects in cold-atom systems.
These are relevant in view of recent developments in theoretical and experimental approaches in the field, see, e.g., Refs.~\cite{Mordini2020,Bulakhov2021}.

\begin{acknowledgments}
The authors acknowledge support by the National Research Foundation of Ukraine, Grant No.~0120U104963 and the Ministry of Education and Science of Ukraine, Research Grant No.~0120U102252.
Access to computing and storage facilities provided by the Poznan Supercomputing and Networking Center (EAGLE cluster) is greatly appreciated.
\end{acknowledgments}

\bibliography{A42}	

\begin{thebibliography}{20}%
\makeatletter
\providecommand \@ifxundefined [1]{%
 \@ifx{#1\undefined}
}%
\providecommand \@ifnum [1]{%
 \ifnum #1\expandafter \@firstoftwo
 \else \expandafter \@secondoftwo
 \fi
}%
\providecommand \@ifx [1]{%
 \ifx #1\expandafter \@firstoftwo
 \else \expandafter \@secondoftwo
 \fi
}%
\providecommand \natexlab [1]{#1}%
\providecommand \enquote  [1]{``#1''}%
\providecommand \bibnamefont  [1]{#1}%
\providecommand \bibfnamefont [1]{#1}%
\providecommand \citenamefont [1]{#1}%
\providecommand \href@noop [0]{\@secondoftwo}%
\providecommand \href [0]{\begingroup \@sanitize@url \@href}%
\providecommand \@href[1]{\@@startlink{#1}\@@href}%
\providecommand \@@href[1]{\endgroup#1\@@endlink}%
\providecommand \@sanitize@url [0]{\catcode `\\12\catcode `\$12\catcode
  `\&12\catcode `\#12\catcode `\^12\catcode `\_12\catcode `\%12\relax}%
\providecommand \@@startlink[1]{}%
\providecommand \@@endlink[0]{}%
\providecommand \url  [0]{\begingroup\@sanitize@url \@url }%
\providecommand \@url [1]{\endgroup\@href {#1}{\urlprefix }}%
\providecommand \urlprefix  [0]{URL }%
\providecommand \Eprint [0]{\href }%
\providecommand \doibase [0]{http://dx.doi.org/}%
\providecommand \selectlanguage [0]{\@gobble}%
\providecommand \bibinfo  [0]{\@secondoftwo}%
\providecommand \bibfield  [0]{\@secondoftwo}%
\providecommand \translation [1]{[#1]}%
\providecommand \BibitemOpen [0]{}%
\providecommand \bibitemStop [0]{}%
\providecommand \bibitemNoStop [0]{.\EOS\space}%
\providecommand \EOS [0]{\spacefactor3000\relax}%
\providecommand \BibitemShut  [1]{\csname bibitem#1\endcsname}%
\let\auto@bib@innerbib\@empty
\bibitem [{\citenamefont {Dalfovo}\ \emph {et~al.}(1999)\citenamefont
  {Dalfovo}, \citenamefont {Giorgini}, \citenamefont {Pitaevskii},\ and\
  \citenamefont {Stringari}}]{Dalfovo1999}%
  \BibitemOpen
  \bibfield  {author} {\bibinfo {author} {\bibfnamefont {F.}~\bibnamefont
  {Dalfovo}}, \bibinfo {author} {\bibfnamefont {S.}~\bibnamefont {Giorgini}},
  \bibinfo {author} {\bibfnamefont {L.~P.}\ \bibnamefont {Pitaevskii}}, \ and\
  \bibinfo {author} {\bibfnamefont {S.}~\bibnamefont {Stringari}},\ }\href
  {\doibase 10.1103/RevModPhys.71.463} {\bibfield  {journal} {\bibinfo
  {journal} {Rev. Mod. Phys.}\ }\textbf {\bibinfo {volume} {71}},\ \bibinfo
  {pages} {463} (\bibinfo {year} {1999})}\BibitemShut {NoStop}%
\bibitem [{\citenamefont {Giorgini}\ \emph {et~al.}(2008)\citenamefont
  {Giorgini}, \citenamefont {Pitaevskii},\ and\ \citenamefont
  {Stringari}}]{Giorgini2008}%
  \BibitemOpen
  \bibfield  {author} {\bibinfo {author} {\bibfnamefont {S.}~\bibnamefont
  {Giorgini}}, \bibinfo {author} {\bibfnamefont {L.~P.}\ \bibnamefont
  {Pitaevskii}}, \ and\ \bibinfo {author} {\bibfnamefont {S.}~\bibnamefont
  {Stringari}},\ }\href {\doibase 10.1103/RevModPhys.80.1215} {\bibfield
  {journal} {\bibinfo  {journal} {Rev. Mod. Phys.}\ }\textbf {\bibinfo {volume}
  {80}},\ \bibinfo {pages} {1215} (\bibinfo {year} {2008})}\BibitemShut
  {NoStop}%
\bibitem [{\citenamefont {Pethick}\ and\ \citenamefont
  {Smith}(2002)}]{Pethick2002}%
  \BibitemOpen
  \bibfield  {author} {\bibinfo {author} {\bibfnamefont {C.~J.}\ \bibnamefont
  {Pethick}}\ and\ \bibinfo {author} {\bibfnamefont {H.}~\bibnamefont
  {Smith}},\ }\href {\doibase 10.1017/CBO9780511802850} {\emph {\bibinfo
  {title} {Bose-Einstein Condensation in Dilute Gases}}}\ (\bibinfo
  {publisher} {Cambridge University Press},\ \bibinfo {address} {Cambridge},\
  \bibinfo {year} {2002})\BibitemShut {NoStop}%
\bibitem [{\citenamefont {Pitaevskii}\ and\ \citenamefont
  {Stringari}(2003)}]{Pitaevskii2003}%
  \BibitemOpen
  \bibfield  {author} {\bibinfo {author} {\bibfnamefont {L.~P.}\ \bibnamefont
  {Pitaevskii}}\ and\ \bibinfo {author} {\bibfnamefont {S.}~\bibnamefont
  {Stringari}},\ }\href {\doibase 10.1093/acprof:oso/9780198758884.001.0001}
  {\emph {\bibinfo {title} {Bose-Einstein Condensation}}}\ (\bibinfo
  {publisher} {Clarendon Press},\ \bibinfo {address} {Oxford},\ \bibinfo {year}
  {2003})\BibitemShut {NoStop}%
\bibitem [{\citenamefont {Pathria}\ and\ \citenamefont
  {Beale}(2011)}]{Pathria2011}%
  \BibitemOpen
  \bibfield  {author} {\bibinfo {author} {\bibfnamefont {R.~K.}\ \bibnamefont
  {Pathria}}\ and\ \bibinfo {author} {\bibfnamefont {P.~D.}\ \bibnamefont
  {Beale}},\ }\href {\doibase 10.1016/C2009-0-62310-2} {\emph {\bibinfo {title}
  {Statistical mechanics}}}\ (\bibinfo  {publisher} {Elsevier},\ \bibinfo
  {address} {Burlington},\ \bibinfo {year} {2011})\BibitemShut {NoStop}%
\bibitem [{\citenamefont {Truscott}\ \emph {et~al.}(2001)\citenamefont
  {Truscott}, \citenamefont {Strecker}, \citenamefont {McAlexander},
  \citenamefont {Partridge},\ and\ \citenamefont {Hulet}}]{Truscott2001}%
  \BibitemOpen
  \bibfield  {author} {\bibinfo {author} {\bibfnamefont {A.~G.}\ \bibnamefont
  {Truscott}}, \bibinfo {author} {\bibfnamefont {K.~E.}\ \bibnamefont
  {Strecker}}, \bibinfo {author} {\bibfnamefont {W.~I.}\ \bibnamefont
  {McAlexander}}, \bibinfo {author} {\bibfnamefont {G.~B.}\ \bibnamefont
  {Partridge}}, \ and\ \bibinfo {author} {\bibfnamefont {R.~G.}\ \bibnamefont
  {Hulet}},\ }\href {\doibase 10.1126/science.1059318} {\bibfield  {journal}
  {\bibinfo  {journal} {Science}\ }\textbf {\bibinfo {volume} {291}},\ \bibinfo
  {pages} {2570} (\bibinfo {year} {2001})}\BibitemShut {NoStop}%
\bibitem [{\citenamefont {Romero-Roch{\'\i}n}\ and\ \citenamefont
  {Bagnato}(2005)}]{Romero2005}%
  \BibitemOpen
  \bibfield  {author} {\bibinfo {author} {\bibfnamefont {V.}~\bibnamefont
  {Romero-Roch{\'\i}n}}\ and\ \bibinfo {author} {\bibfnamefont {V.~S.}\
  \bibnamefont {Bagnato}},\ }\href@noop {} {\bibfield  {journal} {\bibinfo
  {journal} {Braz. J. Phys.}\ }\textbf {\bibinfo {volume} {35}},\ \bibinfo
  {pages} {607} (\bibinfo {year} {2005})}\BibitemShut {NoStop}%
\bibitem [{\citenamefont {Sotnikov}\ \emph {et~al.}(2017)\citenamefont
  {Sotnikov}, \citenamefont {Sereda},\ and\ \citenamefont
  {Slyusarenko}}]{Sotnikov2017LTP}%
  \BibitemOpen
  \bibfield  {author} {\bibinfo {author} {\bibfnamefont {A.~G.}\ \bibnamefont
  {Sotnikov}}, \bibinfo {author} {\bibfnamefont {K.~V.}\ \bibnamefont
  {Sereda}}, \ and\ \bibinfo {author} {\bibfnamefont {Y.~V.}\ \bibnamefont
  {Slyusarenko}},\ }\href {\doibase 10.1063/1.4975807} {\bibfield  {journal}
  {\bibinfo  {journal} {Low Temp. Phys.}\ }\textbf {\bibinfo {volume} {43}},\
  \bibinfo {pages} {144} (\bibinfo {year} {2017})}\BibitemShut {NoStop}%
\bibitem [{\citenamefont {Ensher}\ \emph {et~al.}(1996)\citenamefont {Ensher},
  \citenamefont {Jin}, \citenamefont {Matthews}, \citenamefont {Wieman},\ and\
  \citenamefont {Cornell}}]{Ensher1996}%
  \BibitemOpen
  \bibfield  {author} {\bibinfo {author} {\bibfnamefont {J.~R.}\ \bibnamefont
  {Ensher}}, \bibinfo {author} {\bibfnamefont {D.~S.}\ \bibnamefont {Jin}},
  \bibinfo {author} {\bibfnamefont {M.~R.}\ \bibnamefont {Matthews}}, \bibinfo
  {author} {\bibfnamefont {C.~E.}\ \bibnamefont {Wieman}}, \ and\ \bibinfo
  {author} {\bibfnamefont {E.~A.}\ \bibnamefont {Cornell}},\ }\href {\doibase
  10.1103/PhysRevLett.77.4984} {\bibfield  {journal} {\bibinfo  {journal}
  {Phys. Rev. Lett.}\ }\textbf {\bibinfo {volume} {77}},\ \bibinfo {pages}
  {4984} (\bibinfo {year} {1996})}\BibitemShut {NoStop}%
\bibitem [{Note1()}]{Note1}%
  \BibitemOpen
  \bibinfo {note} {In Fig.~\ref {fig:3_color_dens} the colormaps are given in
  arbitrary units; for quantitative analysis see Fig.~\ref
  {fig:nx_nz}}\BibitemShut {NoStop}%
\bibitem [{Note2()}]{Note2}%
  \BibitemOpen
  \bibinfo {note} {Exact numerical calculation yields $N_0\approx 11$ at
  $T=T_{\protect \rm c}^{\protect \rm sc}$ for the chosen set of trap
  parameters and $N=22000$.}\BibitemShut {Stop}%
\bibitem [{\citenamefont {Grossmann}\ and\ \citenamefont
  {Holthaus}(1995)}]{Grossmann1995}%
  \BibitemOpen
  \bibfield  {author} {\bibinfo {author} {\bibfnamefont {S.}~\bibnamefont
  {Grossmann}}\ and\ \bibinfo {author} {\bibfnamefont {M.}~\bibnamefont
  {Holthaus}},\ }\href {\doibase https://doi.org/10.1016/0375-9601(95)00766-V}
  {\bibfield  {journal} {\bibinfo  {journal} {Phys. Lett. A}\ }\textbf
  {\bibinfo {volume} {208}},\ \bibinfo {pages} {188} (\bibinfo {year}
  {1995})}\BibitemShut {NoStop}%
\bibitem [{\citenamefont {Jaouadi}\ \emph {et~al.}(2011)\citenamefont
  {Jaouadi}, \citenamefont {Telmini},\ and\ \citenamefont
  {Charron}}]{Jaouadi2011}%
  \BibitemOpen
  \bibfield  {author} {\bibinfo {author} {\bibfnamefont {A.}~\bibnamefont
  {Jaouadi}}, \bibinfo {author} {\bibfnamefont {M.}~\bibnamefont {Telmini}}, \
  and\ \bibinfo {author} {\bibfnamefont {E.}~\bibnamefont {Charron}},\ }\href
  {\doibase 10.1103/PhysRevA.83.023616} {\bibfield  {journal} {\bibinfo
  {journal} {Phys. Rev. A}\ }\textbf {\bibinfo {volume} {83}},\ \bibinfo
  {pages} {023616} (\bibinfo {year} {2011})}\BibitemShut {NoStop}%
\bibitem [{Note3()}]{Note3}%
  \BibitemOpen
  \bibinfo {note} {At $N<100$ the temperature dependencies of the specific heat
  cease to have a vivid maximum associated with the changes in the many-body
  state.}\BibitemShut {Stop}%
\bibitem [{\citenamefont {Griesmaier}\ \emph {et~al.}(2005)\citenamefont
  {Griesmaier}, \citenamefont {Werner}, \citenamefont {Hensler}, \citenamefont
  {Stuhler},\ and\ \citenamefont {Pfau}}]{Griesmaier2005}%
  \BibitemOpen
  \bibfield  {author} {\bibinfo {author} {\bibfnamefont {A.}~\bibnamefont
  {Griesmaier}}, \bibinfo {author} {\bibfnamefont {J.}~\bibnamefont {Werner}},
  \bibinfo {author} {\bibfnamefont {S.}~\bibnamefont {Hensler}}, \bibinfo
  {author} {\bibfnamefont {J.}~\bibnamefont {Stuhler}}, \ and\ \bibinfo
  {author} {\bibfnamefont {T.}~\bibnamefont {Pfau}},\ }\href {\doibase
  10.1103/PhysRevLett.94.160401} {\bibfield  {journal} {\bibinfo  {journal}
  {Phys. Rev. Lett.}\ }\textbf {\bibinfo {volume} {94}},\ \bibinfo {pages}
  {160401} (\bibinfo {year} {2005})}\BibitemShut {NoStop}%
\bibitem [{\citenamefont {Ketterle}\ and\ \citenamefont {van
  Druten}(1996)}]{Ketterle1996}%
  \BibitemOpen
  \bibfield  {author} {\bibinfo {author} {\bibfnamefont {W.}~\bibnamefont
  {Ketterle}}\ and\ \bibinfo {author} {\bibfnamefont {N.~J.}\ \bibnamefont {van
  Druten}},\ }\href {\doibase 10.1103/PhysRevA.54.656} {\bibfield  {journal}
  {\bibinfo  {journal} {Phys. Rev. A}\ }\textbf {\bibinfo {volume} {54}},\
  \bibinfo {pages} {656} (\bibinfo {year} {1996})}\BibitemShut {NoStop}%
\bibitem [{\citenamefont {Yukalov}(2005)}]{Yukalov2005}%
  \BibitemOpen
  \bibfield  {author} {\bibinfo {author} {\bibfnamefont {V.~I.}\ \bibnamefont
  {Yukalov}},\ }\href {\doibase 10.1103/PhysRevA.72.033608} {\bibfield
  {journal} {\bibinfo  {journal} {Phys. Rev. A}\ }\textbf {\bibinfo {volume}
  {72}},\ \bibinfo {pages} {033608} (\bibinfo {year} {2005})}\BibitemShut
  {NoStop}%
\bibitem [{\citenamefont {Giorgini}\ \emph {et~al.}(1996)\citenamefont
  {Giorgini}, \citenamefont {Pitaevskii},\ and\ \citenamefont
  {Stringari}}]{Giorgini1996}%
  \BibitemOpen
  \bibfield  {author} {\bibinfo {author} {\bibfnamefont {S.}~\bibnamefont
  {Giorgini}}, \bibinfo {author} {\bibfnamefont {L.~P.}\ \bibnamefont
  {Pitaevskii}}, \ and\ \bibinfo {author} {\bibfnamefont {S.}~\bibnamefont
  {Stringari}},\ }\href {\doibase 10.1103/PhysRevA.54.R4633} {\bibfield
  {journal} {\bibinfo  {journal} {Phys. Rev. A}\ }\textbf {\bibinfo {volume}
  {54}},\ \bibinfo {pages} {R4633} (\bibinfo {year} {1996})}\BibitemShut
  {NoStop}%
\bibitem [{\citenamefont {Mordini}\ \emph {et~al.}(2020)\citenamefont
  {Mordini}, \citenamefont {Trypogeorgos}, \citenamefont {Farolfi},
  \citenamefont {Wolswijk}, \citenamefont {Stringari}, \citenamefont
  {Lamporesi},\ and\ \citenamefont {Ferrari}}]{Mordini2020}%
  \BibitemOpen
  \bibfield  {author} {\bibinfo {author} {\bibfnamefont {C.}~\bibnamefont
  {Mordini}}, \bibinfo {author} {\bibfnamefont {D.}~\bibnamefont
  {Trypogeorgos}}, \bibinfo {author} {\bibfnamefont {A.}~\bibnamefont
  {Farolfi}}, \bibinfo {author} {\bibfnamefont {L.}~\bibnamefont {Wolswijk}},
  \bibinfo {author} {\bibfnamefont {S.}~\bibnamefont {Stringari}}, \bibinfo
  {author} {\bibfnamefont {G.}~\bibnamefont {Lamporesi}}, \ and\ \bibinfo
  {author} {\bibfnamefont {G.}~\bibnamefont {Ferrari}},\ }\href {\doibase
  10.1103/PhysRevLett.125.150404} {\bibfield  {journal} {\bibinfo  {journal}
  {Phys. Rev. Lett.}\ }\textbf {\bibinfo {volume} {125}},\ \bibinfo {pages}
  {150404} (\bibinfo {year} {2020})}\BibitemShut {NoStop}%
\bibitem [{\citenamefont {Bulakhov}\ \emph {et~al.}(2021)\citenamefont
  {Bulakhov}, \citenamefont {Peletminskii}, \citenamefont {Slyusarenko},\ and\
  \citenamefont {Sotnikov}}]{Bulakhov2021}%
  \BibitemOpen
  \bibfield  {author} {\bibinfo {author} {\bibfnamefont {M.~S.}\ \bibnamefont
  {Bulakhov}}, \bibinfo {author} {\bibfnamefont {A.~S.}\ \bibnamefont
  {Peletminskii}}, \bibinfo {author} {\bibfnamefont {Y.~V.}\ \bibnamefont
  {Slyusarenko}}, \ and\ \bibinfo {author} {\bibfnamefont {A.~G.}\ \bibnamefont
  {Sotnikov}},\ }\href {\doibase 10.1088/1402-4896/abdcf5} {\bibfield
  {journal} {\bibinfo  {journal} {Phys. Scr.}\ }\textbf {\bibinfo {volume}
  {96}},\ \bibinfo {pages} {045401} (\bibinfo {year} {2021})}\BibitemShut
  {NoStop}%
\end{thebibliography}%
\end{document}